\documentclass{article}
\usepackage{spconf,amsmath,graphicx,booktabs,xcolor,multirow,enumitem}

\title{COUGH ACTIVITY DETECTION FOR AUTOMATIC TUBERCULOSIS SCREENING}
\name{\vspace{-10mm}}

\address{
    \textit{Joshua Jansen van V\"{u}ren$^1$, Devendra Singh Parihar$^1$, Daphne Naidoo$^2$, Kimsey Zajac$^3$}, \\
    \textit{Willy Ssengooba$^4$, Grant Theron$^2$, Thomas Niesler$^1$}
    \thanks{We acknowledge funding from the EDCTP2 program supported by the European Union (RIA2020I-3305, CAGE-TB). This project is also supported by Global Health EDCTP3 and its members (101145817, 4-CAGE-TB).}\\\\
\normalsize
$^1$Department of Electrical and Electronic Engineering, Stellenbosch University, Stellenbosch, South Africa \\[0.1cm]
\normalsize
$^2$DSI-NRF Centre of Excellence for Biomedical Tuberculosis Research,\\[-0.1cm] \normalsize South African Medical Research Council Centre for Tuberculosis Research,\\[-0.1cm] \normalsize Division of Molecular Biology and Human Genetics,\\[-0.1cm] \normalsize Faculty of Medicine and Health Sciences, Stellenbosch University, Cape Town, South Africa \\[0.1cm]
\normalsize
$^3$Faculty of Business and Economics, University of G\"{o}ttingen, G\"{o}ttingen, Germany \\[0.1cm]
\normalsize
$^4$Department of Medical Microbiology, College of Health Sciences, Makerere University,\\[-0.1cm]\normalsize Biomedical Research Center (MAKBRC), Kampala, Uganda
}

\begin{document}
\maketitle
\begin{abstract}
The automatic identification of cough segments in audio through the determination of start and end points is pivotal to building scalable screening tools in health technologies for pulmonary related diseases.
We propose the application of two current pre-trained architectures to the task of cough activity detection.
A dataset of recordings containing cough from patients symptomatic for tuberculosis (TB) who self-present at community-level care centres in South Africa and Uganda is employed.
When automatic start and end points are determined using XLS-R, an average precision of 0.96 and an area under the receiver-operating characteristic of 0.99 are achieved for the test set.
We show that best average precision is achieved by utilising only the first three layers of the network, which has the dual benefits of reduced computational and memory requirements, pivotal for smartphone-based applications.
This XLS-R configuration is shown to outperform an audio spectrogram transformer (AST) as well as a logistic regression baseline by 9\% and 27\% absolute in test set average precision respectively.
Furthermore, a downstream TB classification model trained using the coughs automatically isolated by XLS-R comfortably outperforms a model trained on the coughs isolated by AST, and is only narrowly outperformed by a classifier trained on the ground truth coughs.
We conclude that the application of large pre-trained transformer models is an effective approach to identifying cough end-points and that the integration of such a model into a screening tool is feasible.
\end{abstract}
\section{Introduction}
\label{sec:intro}

The collection of cough sounds for use in mobile health (mHealth) technologies is an area of growing interest~\cite{cough_review_praveer,trends_in_cough_review}.
Several studies have collected cough sounds through online crowdsourcing or in clinical settings.
These have been applied to classify various pulmonary related diseases, such as pneumonia~\cite{child_bronchitis_pneumonia_cough_class}, coronavirus disease~\cite{sharma2020coswara,coughvid2021,pizzo2021iatosaipoweredprescreeningtool}, TB~\cite{tbscreen,Huddart2024}, chronic obstructive pulmonary disease~\cite{healthinf18,PorterPA4278}, and asthma~\cite{7311223}.
Predominately, recordings containing coughs are manually annotated. 
Although previous work has utilised machine learning based methods to identify coughs within an audio signal, to our knowledge no studies have considered the impact of such automatic cough detection on downstream disease classification~\cite{cough_review_praveer}.

In a clinical setting, the manual annotation of coughs may be infeasible due to the time and effort required and additional concerns regarding hygiene.
Therefore, the ability to automatically obtain the start and end points of coughs from a recording is attractive, as it allows such manual intervention to be sidestepped, and can be included in the pipeline of an audio-based disease screening tool.
In this scenario, accurate isolation of coughs from audio recordings is of utmost importance as non-cough signals contaminating the input may produce adverse downstream effects.
This work explicitly compares the effect of passing automatically extracted coughs to a TB classification model to providing the same model with input coughs annotated by a human.

Several authors have considered the identification of coughs within audio signals.
An area under the receiver-operating characteristic (AUC) of $96.4 \pm 3.3 \%$ is achieved in \cite{coughvid2021} when tasked to classify if a cough exists within an audio recording.
This is accomplished by employing a XGBoost based classification model which receives as input 68 distinct features such as mel-frequency cepstral coefficients and power spectral densities and zero crossing rate. 
However, no activity detection experiments are conducted.
A frame-wise AUC of $90.7\%$ is achieved when employing SVM to classify cough versus non-cough frames when fed as input the moving average and standard deviation of several spectral based features (centroid, flatness, bandwidth and more)~\cite{8584081}.
A root-mean squared error based cough isolation strategy has been proposed, which was employed to extract coughs from an open source corpus, however because no ground truth manual annotations exist, performance of this classifier is uncertain~\cite{ashby2022novel}.
In~\cite{7570164} a convolutional neural network (CNN) is compared to a encoder-decoder recurrent neural network (RNN) for cough and speech discrimination.
Competitive performance is found between the two models, achieving frame-wise AUCs for $95\%$ and $96\%$ respectively. 
Using SincNET~\cite{sincnet}, a framewise AUC of $98.7\%$ is achieved in~\cite{SHARAN2023104580} on a subset of 100 manually annotated recordings from CoughVID~\cite{coughvid2021}.

The application of transformer-based models to the task of cough activity detection from audio recordings has not been investigated. 
We consider the application of XLS-R~\cite{babu22_interspeech}, a 300 million parameter transformer which has been trained on over 400,000 hours of speech in over 128 languages in a self-supervised manner.
We compare the performance of XLS-R with a well established spectrogram-based transformer, the AST~\cite{gong21b_interspeech}.
This model has been pretrained on a substantially different corpus of general sounds~\cite{audioset} and not solely on speech as for XLS-R.
This serves as an interesting comparison as these models differ both in terms of pretraining data as well as the strategies employed for feature extraction.
Although transformers are generally more computationally expensive than classical machine learning or domain specific deep learning models, this is typically contrasted by significant performance gains.
To verify this hypothesis, and as a further baseline, we consider a logistic regression (LR) model configured to be reminiscent of a time-delay neural network~\cite{tdnn}.

The aim of this paper is therefore two fold.
Firstly, we aim to determine if transformer-based networks are effective at cough activity detection and contrast their performance to a baseline model.
Secondly, we investigate whether using coughs automatically isolated degrades TB classification performance compared to manual annotations.

The remainder of the paper is structured as follows.
Section~\ref{sec:data} presents the dataset employed in this study.
Section~\ref{sec:experimental_setup} describes the experimental setup, from which Section~\ref{sec:results} presents the cough activity detection as well as the TB classification results.
Finally, Section~\ref{sec:conclusion} concludes.

\section{Data}
\label{sec:data}

This work utilises cough recordings collected from 1193 patients in both South Africa and Uganda as part of a larger project which aims to identify people at risk for TB who can subsequently be referred for confirmatory testing using only cough sounds (CAGE-TB, 4-CAGE-TB).

The dataset contains 464 recordings from patients in South Africa, collected from three different community-level care centres, and 729 patients in Uganda, also collected from three community-level care centres.
The recording procedure begins with the health care worker stating the patient identification number as well as the current date and time for record keeping and later quality control.
After this, the patient is asked to count from one to ten.
Finally, the patient is asked to cough a number of times, after which the health care worker states that the recording is complete.
For all recordings, manual annotations of both the counting and the coughing were produced utilising the ELAN~\cite{wittenburg_elan} software package.
In total, the dataset contains 21,808 end-pointed coughs, which amounts to 2.52 hours of cough audio, and 20.2 hours of other audio.
Over the dataset, the mean and standard deviation of the cough length is $416 \pm 207$ milliseconds. 

It is important to note that, because these recordings are collected in busy primary health care facilities, often outside, background noise produced by for example motor vehicles, construction, or generators, was common.
Annotators were instructed only to disregard coughs when they were not perceivable over the background noise.
By utilising these annotations as ground truth targets, we train a neural network to detect cough sounds which are well distinguished from background noise.
Additionally, it is crucial that these systems are evaluated settings such as these, as they are representative of where TB triage or screening will take place in practice.

\begin{figure}
    \centering
    \includegraphics[width=\linewidth]{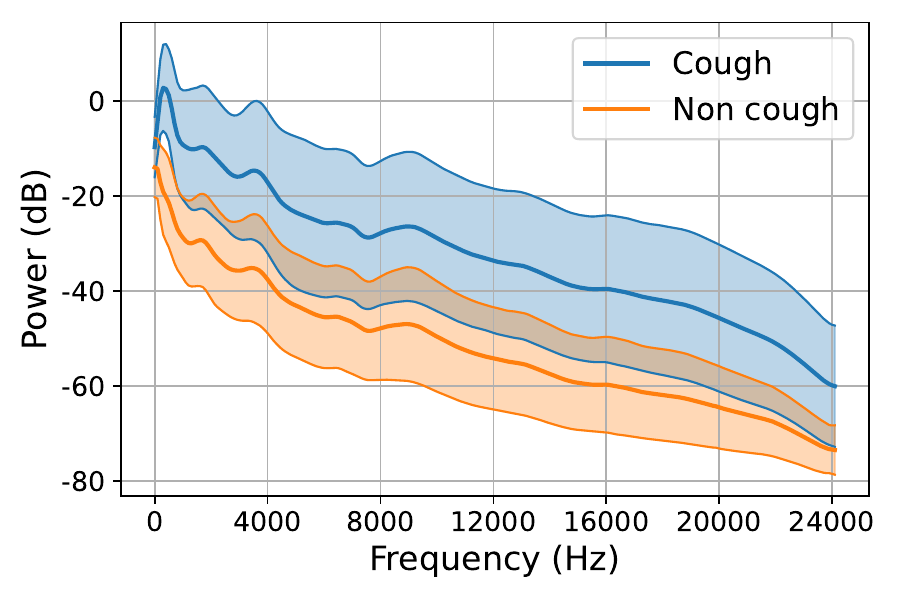}
    \caption{Power (dB) versus frequency (Hz) for all cough (blue) and non-cough (orange) audio, shaded regions indicate standard deviation over corpus.}
    \label{fig:pow_vs_freq_cage}
\end{figure}

In Figure~\ref{fig:pow_vs_freq_cage} the average power versus frequency is presented for all cough and non-cough audio.
It is clear that, in this dataset, the coughs are on average of higher power than other signal components.
Importantly, the majority of the cough power is present below 8kHz.
Therefore, the application of pretrained transformer models which are limited to a fixed sampling frequency of 16kHz is feasible.

We partition the data into training, development and test sets, as presented in Table~\ref{tab:dataset_statistics}.
The training and development data is drawn exclusively from the Ugandan sub-cohort, with approximately 75\% included for training and the remainder for development. 
The test set comprises exclusively data drawn from the South African sub-cohort.
The premise for this selection is two fold.
Firstly because it presents a challenging test condition, as the 
recordings locations are unique across the sets, and therefore pose novel environmental conditions.
Additionally, the languages spoken in the recordings differ, English and Luganda occurs in the Ugandan recordings, while Afrikaans, English and isiXhosa occurs in the South African recordings.
The second reason for this selection is that the cough detector can be trained and optimised on the Ugandan sub-cohort and then applied to the South African test sub-cohort.
The automatic start and end points collectively form a dataset which contains automatically isolated coughs.
This dataset can then be used to train TB classification models.
In this way the effect of different cough activity detection models on TB classification performance can be evaluated.

\begin{table}[]
    \caption{Dataset training, development and test partition statistics. The training and development sets comprise only data collected in Uganda, while the testing data comprises only recordings collected in South Africa.}
    \vspace{2mm}
    \label{tab:dataset_statistics}
    \centering
    \begin{tabular}{c c c c c}
        \toprule
        \multirow{2}{*}{Part} & Number     & Duration       & Number & Duration \\
             & recordings & recordings     & coughs & coughs \\
        \midrule
        Train   & 583   & 13.20 & 11890 & 1.37 \\
        Dev     & 146   & 3.31  & 3015  & 0.36 \\
        Test    & 464   & 7.20  & 6874  & 0.79 \\
        \midrule
        Total   & 1193 & 23.71  & 21791 & 2.52 \\
         \bottomrule
    \end{tabular}
    
\end{table}

\section{Experimental setup}
\label{sec:experimental_setup}

\begin{figure}[h!]
    \centering
    \includegraphics[width=0.9\linewidth]{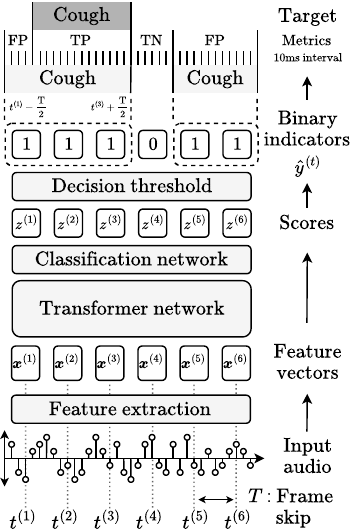}
    \caption{Cough activity detection model architecture and overview of the determination strategy for cough start and end points. TP: True positive, FP: False positive, TN: True negative.}
    \label{fig:network_diagram}
\end{figure}

The cough activity detection models are trained to predict the presence of coughs in a recording on a frame-by-frame basis.
As shown in Figure~\ref{fig:network_diagram}, each model (AST, XLS-R, and LR) applies a feature extraction step to the input audio, which results in a series of feature vectors $\boldsymbol{x}^{(n)}$ each based on an input frame which is centred around timestep $t^{(n)}$ and with frameskip $T$.
Based on these feature vectors, the classifier produces a per-frame scalar score $z^{(i)}$.
These output scores are transformed by a sigmoid function to estimate the posterior probability, denoted as $\hat{y}^{(t)}$, of the presence of cough in the portion of audio delimited by start and end time boundaries $[t^{(n)}-\frac{T}{2}$ and $t^{(n)}+\frac{T}{2}]$ respectively.
The labels used for model training are determined from the manual annotations at the associated timesteps, where the label for each frame is determined by the class which occupies the majority ($> 50\%$) of the frame duration.

Cough start and end times are determined by converting the per-frame scores to binary indicators through the application of a decision threshold.
A sequence of consecutive frames positively identified as containing cough is then labelled as a cough, as illustrated in Figure~\ref{fig:network_diagram}.
In this example, the cough start and end-times are the start of the first frame ($t^{(1)} - \frac{T}{2}$) and end of the last frame ($t^{(3)} +\frac{T}{2}$) respectively.
In subsequent experiments, median filtering is additionally applied to this sequence of binary indicators.
No other post-processing, such as minimum cough duration and minimum cough separation is applied.

Since both the AST and XLS-R are pre-trained models, the frame length and skip are fixed and cannot be freely chosen.
For the AST, the default feature extraction strategy utilises a 128-dimensional mel-spaced filterbank.
A series of concatenated vectors extracted from $16 \times 16$ patches of this mel-spectrogram are the inputs to the AST.
Patches are then extracted using a stride of 10 in both time and frequency, this corresponds to a frame length of 160ms and a frame skip of 100ms.
XLS-R directly receives the input waveform, which is fed through a CNN network whose kernel size and stride is designed to produce a frame size of 25ms and a frame skip of 20ms.  
For LR we choose to employ the same patching behaviour as for the AST. 
Therefore, the AST and LR architectures will produce one output classification every 100 ms while XLS-R produces an output every 20ms.

\subsection{Cough activity detection models}

\begin{figure*}[ht]
    \centering
    \includegraphics[width=0.8\linewidth]{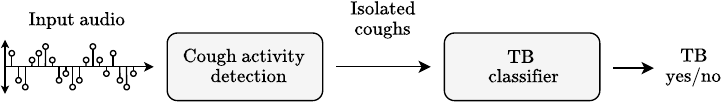}
    \caption{Top level diagram of automatic cough isolation from raw audio to downstream TB classification.}
    \label{fig:top_level_cough_isolation_and_TB_classification}
\end{figure*}

We fine-tune each of the considered transformers on our training set (Table~\ref{tab:dataset_statistics}) utilising the AdamW~\cite{loshchilov2018decoupled} optimiser.
The learning rate is increased linearly from zero for the first ten percent of training steps to a defined maximum, after which it is linearly decreased to zero for the remainder of the training steps.
All models are trained for 16 epochs.

For both transformer networks, a two layer neural network is included after the final transformer layer, in order to produce the posterior cough probability as output.
This network receives as input the hidden state vectors of the transformer models.
For XLS-R, the first layer of the network reduces the dimensionality by half (from $1024$ to $512$) and then applies a GeLU~\cite{hendrycks2023gaussianerrorlinearunits} activation function.
The final layer receives the resulting $512$ dimensional features and produces a single binary output, which is trained to indicate whether a cough is detected in the frame or not.

As for XLS-R, 
the $768$ dimensional output vectors computed by the AST are transformed to $384$-dimensional vectors by the first of the two layer network, after which a GeLU activation function is applied.
Considering the striding across the frequency dimension, $\lfloor \frac{128 - 16}{10} \rfloor + 1 = 12$ output patches are extracted from partially-overlapping frequency content at the same 16 frames in the mel-spectrogram.
The corresponding $12$ vectors at the output of the first layer of the final 2-layer network are then concatenated to yield a vector with $4608$ dimensions, which is the input to the final linear layer to again produce a single binary output.

Finally, as a baseline model, we include LR, which receives as input a portion of the mel-spectrogram as described above and produces as output an estimate of the posterior probability of cough per frame.
This classifier is also trained for 16 epochs however no learning rate scheduler is applied.
\subsection{Evaluation metrics}

Cough activity detection models are evaluated using frame-wise AUC, average precision (AP), as well as coverage (sensitivity) and purity (precision).
Because each model has a different output rate, frame-wise evaluation metrics are calculated at a fixed 10ms interval as shown in Figure~\ref{fig:network_diagram}.
Therefore, for the AST and LR, each output prediction is divided into 16 equally spaced non-overlapping intervals per output prediction which are then scored, and for XLS-R into two intervals.
This was done so that the output rate of the model does not impact the total number of classifications which are scored, thereby making the figures directly comparable.
The small (10ms) value was chosen to ensure that predicted cough boundaries always align with the boundaries of the interval being scored, since all ground truth start and end times are specified to the nearest 10ms. 

Coverage is defined as $\frac{\text{TP}}{\text{P}}$, where $\text{TP}$ (true positives) are the frames correctly identified as cough and $\text{P}$ (positives) are the total number of true cough frames.
Purity is defined as $\frac{\text{TP}}{\text{PP}}$, where $\text{PP} = \text{TP} + \text{FP}$ is the number of frames both correctly and incorrectly identified as cough, $\text{FP}$ is false positives, and $\text{PP}$ is predicted positives.

Average precision is a useful metric for cough activity detection as it directly reflects the proportion of the true coughs the model is able retrieve (through coverage $\frac{\text{TP}}{\text{P}}$) and the proportion of the audio identified as cough which is indeed coughs (through purity $\frac{\text{TP}}{\text{PP}}$) over a range of decision thresholds.
A well performing cough activity detector should be able to retrieve a large proportion of the ground truth cough frames and produce few false positives.
The dataset under study suffers a large class imbalance ($\approx\mkern1mu1\mkern1.5mu{:}\mkern1.5mu10$ positive to negative).
Therefore, false positive rate ($\text{FPR} = \frac{\text{FP}}{\text{N}}$), which is incorporated into the ROC curve and in turn the AUC score, can appear low because the number of negatives ($\text{N}$) in the denominator is large.
By considering average precision, which incorporates purity, emphasis is placed on reducing false positives in relation to the number of true positives.

The downstream TB classification models are evaluated solely in terms of AUC, as this is the most prevalent metric used in diagnostic testing.

\subsection{Hyperparameter search}
\label{subsec:setup_hyperparameter_search}
Hyperparameters for XLS-R and AST were optimised using a grid search over the following values:
\begin{itemize}[topsep=4pt,itemsep=1pt,partopsep=2pt, parsep=2pt]
    \item Batch size: 8, 16, 32
    \item Learning rate: $1\cdot 10^{-4}$, $1\cdot 10^{-5}$
\end{itemize}
For both XLS-R and AST, the transformer layer from which features are extracted was also treated as a hyperparameter.
Each of the 24 available layers was considered, as was obtaining features directly from the CNN feature extractor for XLS-R.

The hyperparameters considered for the linear classifier were:
\begin{itemize}[topsep=4pt,itemsep=1pt,partopsep=2pt, parsep=2pt]
    \item Batch size of 4, 8, 16
    \item Learning rate: $1\cdot 10^{-2}$, $1\cdot 10^{-3}$
\end{itemize}
All hyperparameters are selected according to the best AP on the development set.

\subsection{TB classification models}

As illustrated in Figure~\ref{fig:top_level_cough_isolation_and_TB_classification}, the TB classification models are trained using the isolated coughs identified automatically using the cough activity detection models or, for comparison, using the manually identified coughs.

Previous research has shown that a bidirectional long short-term memory (LSTM) network is effective when applied to TB cough classification~\cite{frost22_interspeech}.
We consider only this architecture, to ensure that observed differences in TB classification performance can reasonably be attributed to the differences in the cough activity detection. 
The LSTM is optimised over the following hyperparameters: 
\begin{itemize}[topsep=4pt,itemsep=1pt,partopsep=2pt, parsep=2pt]
    \item Batch size: 8, 16, 32
    \item Learning rate: $1\cdot 10^{-4}$, $1\cdot 10^{-5}$ 
    \item Hidden vector dimensionality: 32, 64 
    \item Number of LSTM layers: 1, 2
    \item Dropout: 0.2, 0.6
\end{itemize}
For TB classification, the South African cough data is partitioned into 10 independent folds which each reflect the overall TB prevalence (approximately $18.5\%$).
Hyperparameters are selected independently for each of these folds according to the average development set loss.
To further ensure consistency of the downstream TB classifiers, hyperparameters are selected using the manually annotated dataset.
These hyperparameters are then employed without modification by the classifiers trained on the datasets obtained through automatic cough detection.

\section{Results}
\label{sec:results}

In the following section (Section~\ref{subsec:results_segementation}) we report firstly the cross cohort cough activity detection results.
Section~\ref{subsec:results_segmented_dataset} then discusses the properties of the datasets obtained by applying cough activity detection to the complete recordings available for each patient and compares these to the manual labels.
Section~\ref{subsec:results_median_filtering} presents further results after the application of median filtering.
Finally, Section~\ref{subsec:results_classification} reports the performance of the downstream TB classifiers when trained on either the automatic or the ground truth isolated coughs.

\subsection{Cough activity detection}
\label{subsec:results_segementation}

\begin{figure*}[ht!]
    \centering
    \includegraphics[width=\linewidth]{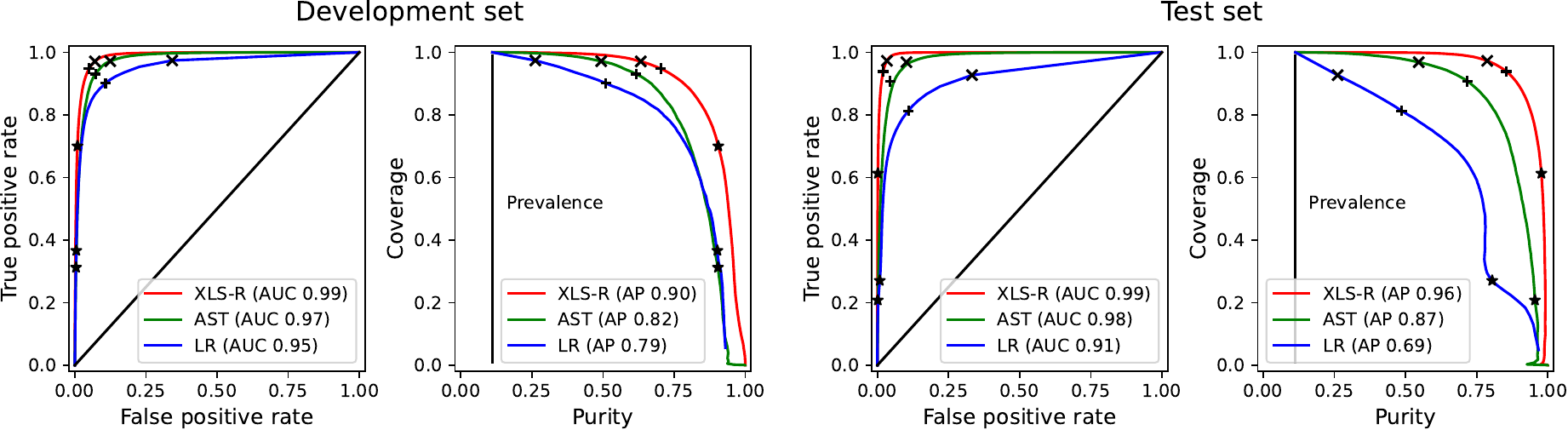}
    \caption{Receiver-operator characteristic and coverage (sensitivity) and purity (precision) curves for the development set for each of the three (XLS-R, AST, LR) cough detectors and the three respective operating points, $\times$: 97\% coverage, $\star$: 90\% purity, and $+$: equal error rate.}
    \label{fig:roc_ap_curves}
\end{figure*}

Table~\ref{tab:optimal_hyperparameters} reports the hyperparameter values obtained using the search procedure outlined in Section~\ref{subsec:setup_hyperparameter_search}.
This table shows that batch size is a stable hyperparameter across models and that a smaller value results in consistently better performance.
Additionally, for both transformer-based models, earlier layers (1 for AST, and 3 for XLS-R) afford better performance in terms of average precision.
This is encouraging, because it means that the subsequent layers in the network can be discarded and model size can be drastically reduced, which is an advantage when considering the potential for deployment on a mobile device.
In the case of XLS-R this reduces model size by a factor of six and improves throughput with a speedup of $3.82\ [3.82,3.83]$ (Mean over 10,000 reruns on a Nvidia A4000 RTX with 95\% confidence interval).

\begin{table}[h!]
    \centering
    \caption{Best performing hyperparameters for the three models (LR, AST, XLS-R).}
    \vspace{2mm}
    \label{tab:optimal_hyperparameters}
    \begin{tabular}{c c c c c c}
        \toprule
         \multirow{2}{*}{Model} & \multirow{2}{*}{Batch} & \multirow{2}{*}{LR} & \multirow{2}{*}{Layer} & \multicolumn{2}{c}{Parameter count} \\
                                & & & & Full & Optimal \\
         \midrule
         LR                 & 8     & 0.01      & - & 2k   & - \\
         AST                & 8     & 0.0001    & 1 & 87M  & 9.5M \\
         XLS-R              & 8     & 0.0001    & 3 & 316M & 51M \\
         \bottomrule
    \end{tabular}
\end{table}

Table~\ref{tab:results_segmentation} presents the development and test set AUC and AP for these optimised models and Figure~\ref{fig:roc_ap_curves} presents the receiver-operator characteristic as well as coverage and purity curves.
From Table~\ref{tab:results_segmentation} it is clear that XLS-R consistently achieves the best performance according to both AUC and average precision for both the development and test sets, outperforming AST by 8\% absolute and LR by 11\% in terms of average precision on the development set and by 9\% (AST) and 27\% (LR) absolute in terms of average precision on the test set.
Although LR is an attractive choice due to its computational efficiency, it appears to not generalise well to the test conditions, suffering a 10\% deterioration in average precision from to the development to the test set. 
This is also reflected in the coverage and purity curves.
The transformer models do not suffer from such degradation in performance on the unseen test data.

\begin{table}[h!]
    \centering
    \caption{Average development and test set performance (AUC, AP) for all three considered models (LR, AST, XLS-R).}
    \vspace{2mm}
    \label{tab:results_segmentation}
    \begin{tabular}{c c c c c c c c}
        \toprule
         \multirow{2}{*}{Model} & \multicolumn{2}{c}{Development} & \multicolumn{2}{c}{Test} \\
          & AUC & AP & AUC & AP \\
         \midrule 
         LR                 & 0.95 & 0.79               & 0.91 & 0.69 \\
         AST                & 0.97 & 0.82               & 0.98 & 0.87 \\
         XLS-R              & \bf{0.99} & \bf{0.90}     & \bf{0.99} & \bf{0.96} \\
         \bottomrule
    \end{tabular}
\end{table}

\subsection{Automatically isolated coughs}
\label{subsec:results_segmented_dataset}

Three thresholds, determined solely on the development set, are employed for each of the three cough activity detection models. This results in nine datasets of automatically extracted coughs. The thresholds were chosen to achieve:
\begin{enumerate}[topsep=4pt,itemsep=1pt,partopsep=2pt, parsep=2pt]
    \item[C :] High coverage (97\%), and therefore sacrifice purity.
    \item[EE:] Equal error rate on the ROC curve.
    \item[P :] High purity (90\%), and therefore sacrifice coverage.
\end{enumerate}
These operating points are indicated for both the development and test sets in Figure~\ref{fig:roc_ap_curves}.

\begin{figure*}
    \centering
    \includegraphics[width=\linewidth]{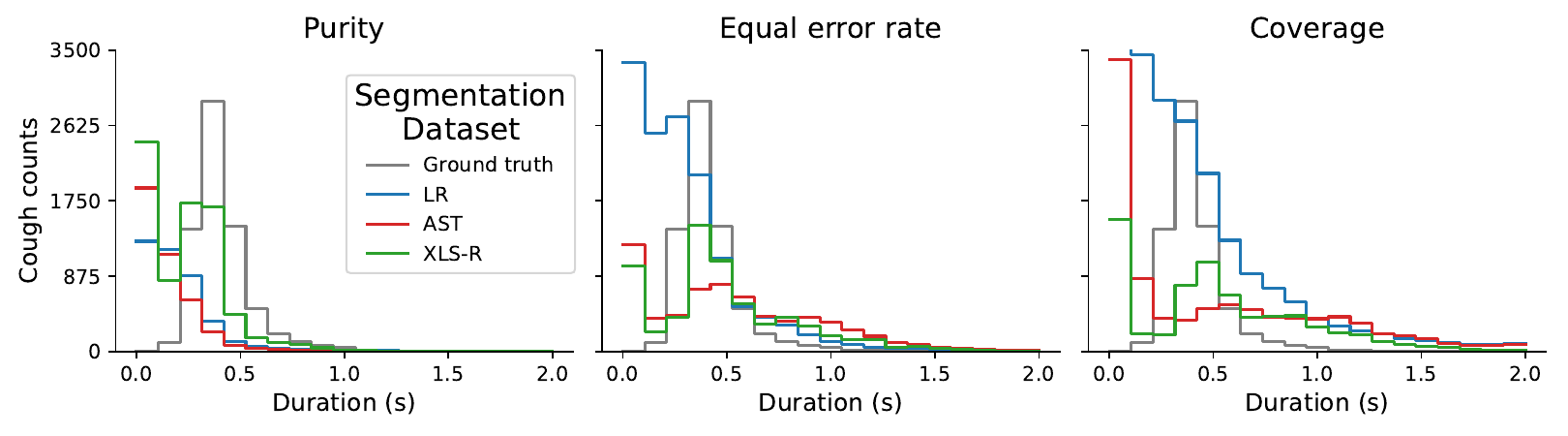}
    \caption{Histogram of test set cough durations for three cough activity detection models (AST, XLS-R, LR) at three different operating points (purity, coverage, equal error rate).}
    \label{fig:automatically_segment_dataset_histogram}
\end{figure*}

In Figure~\ref{fig:automatically_segment_dataset_histogram}, a histogram of cough duration for each of the nine datasets obtained by isolating coughs at each of the three defined operating points (C, EE, P) using each of the three cough activity detection models (LR, AST, XLS-R) is presented.
From the figure it is clear that LR is unable to preserve the distribution of the cough durations at any operating point.
This is especially apparent for the operating point optimised for purity, for which too few coughs are extracted to constitute a dataset which can be used for TB classification. 
In addition, all three models appear to produce fragmented coughs over all three operating points.
This is evidenced by an abnormally high number of coughs lasting  0.1 seconds, as compared to the ground truth distribution.
Table~\ref{tab:automatic_segmented_dataset_stats} presents the statistics for the datasets obtained using the cough detection models.

\begin{table}[h!]
    \centering
    \caption{Statistics of the South African test set for each of the cough datasets obtained using the cough detection models as well as the manually annotated (ground truth) statistics.
    Each of the three cough detection models are used to isolate coughs at three operating points (C: coverage, EE: equal error rate, P: purity) \textdagger: After application of median filter.}
    \vspace{2mm}
    \setlength{\tabcolsep}{4pt}
    \label{tab:automatic_segmented_dataset_stats}
    \begin{tabular}{l l c c c}
        \toprule
        Operating & \multirow{2}{*}{Model} & Number & Average cough & Total dur. \\
        point     &       & coughs & dur. (ms) & (min) \\
        \midrule
        \multicolumn{2}{c}{Ground truth}            & 6886    & 413 $\pm$ 148 & 47.36 \\
        \midrule
                                          & LR      & 22322    & 461 $\pm$ 1623 & 171.58 \\ 
        C                                 & AST     & 9431    & 561 $\pm$ 564 & 88.16 \\ 
                                          & XLS-R   & 6682    & 538 $\pm$ 417 & 59.97 \\ 
        \cmidrule{2-5}
        \multirow{2}{*}{C\textsuperscript{\textdagger}}   
                                          & AST     & 5703    & 903 $\pm$ 738 & 85.82 \\ 
                                          & XLS-R   & 5144    & 697 $\pm$ 380 & 59.73 \\
        \midrule
                                          & LR      & 13585   & 356 $\pm$ 1905 & 80.61 \\ 
        EE                                & AST     & 6487    & 573 $\pm$ 407 & 61.98 \\
                                          & XLS-R   & 6401    & 498 $\pm$ 352 & 53.08 \\ 
        \cmidrule{2-5}
        \multirow{2}{*}{EE\textsuperscript{\textdagger}}   
                                          & AST     & 4962    & 723 $\pm$ 405 & 59.83 \\ 
                                          & XLS-R   & 5289    & 602 $\pm$ 337 & 53.06 \\
        \midrule
                                          & LR      & 3988    & 244 $\pm$ 198 & 16.19 \\ 
        P                                 & AST     & 4017    & 194 $\pm$ 122 & 13.00 \\ 
                                          & XLS-R   & 7544    & 241 $\pm$ 188 & 30.32 \\ 
        \bottomrule
    \end{tabular}
\end{table}

\subsection{Median filtering}
\label{subsec:results_median_filtering}

A median filter is applied to the per-frame classification sequences computed by the two transformer-based cough detectors.
The LR model was not investigated further because it failed to extract reasonable cough start and end-points. 
Neither was any automatically-obtained dataset optimised for purity (P) investigated further, as these models sacrifice coverage.
It was found that the application of median filtering, especially for larger filter sizes, further reduced coverage, and therefore, its application is counterproductive in this situation.

Filter lengths of 3, 5, 7, and 9 frames are investigated.
For the AST this corresponds to filter lengths of 300, 500, 700 and 900ms while for XLS-R this corresponds to filter lengths of 60, 100, 140 and 180ms.
We also investigate filter lengths for XLS-R which correspond to 300, 500 and 700ms for a more direct comparison with AST.
Development set AUC and average precision is reported for these median filter sizes in Table~\ref{tab:median_filter_results}.
None of the investigated median filter sizes outperform the raw output for AST, while a median filter of 140ms is found to offer a small improvement for XLS-R.

\begin{table}[h!]
    \setlength{\tabcolsep}{5pt}
    \centering
    \caption{Development set average precision for AST and XLS-R with best hyperparameters (Table~\ref{tab:optimal_hyperparameters}) after further application of a median filter with various sizes.}
    \vspace{2mm}
    \begin{tabular}{c c c c c}
        \toprule
        \multirow{2}{*}{Model} & Median filter & Approx.     & \multicolumn{2}{c}{Development} \\
              & size          & duration (ms)  & AUC & AP \\
        \midrule
        \multirow{5}{*}{AST}       
               & None & - & 0.9732 & \bf{0.8234} \\
               & 3 & 300 & 0.9735 & 0.8223 \\
               & 5 & 500 & 0.9716 & 0.8009 \\
               & 7 & 700 & 0.9675 & 0.7635 \\
               & 9 & 900 & 0.9612 & 0.7279 \\
        \midrule       
        \multirow{8}{*}{XLS-R}       
               & None & - & 0.9870 & 0.9035 \\
               & 3 & 60 & 0.9871 &  0.9043 \\
               & 5 & 100 & 0.9872 &  0.9047 \\
               & 7 & 140 & 0.9872 &  \bf{0.9048} \\
               & 9 & 180 & 0.9872 &  0.9046 \\
               & 15 & 300 & 0.9866 & 0.8986 \\
               & 25 & 500 & 0.9837 & 0.8649 \\
               & 35 & 700 & 0.9777 & 0.8061 \\
        \bottomrule
    \end{tabular}
    \label{tab:median_filter_results}
\end{table}

\subsection{TB classification}
\label{subsec:results_classification}

For a subset of the nine datasets, we determined the downstream TB classification results for a bidirectional LSTM and present the results in Table~\ref{tab:results_lstm_classification_results}.
Again, we exclude the dataset optimised for coverage and the dataset optimised for purity extracted using LR, as these configurations failed to produce a viable dataset, as already described above.
This is evidenced by the large inconsistency with the ground truth distributions in Table~\ref{tab:automatic_segmented_dataset_stats} and Figure~\ref{fig:automatically_segment_dataset_histogram}.

The TB classifier which is trained on the coughs automatically isolated by XLS-R is able to outperform the TB classifiers trained on the automatically isolated coughs produced by the AST and by LR at all three operating points (C, EE, P).
Amongst the cough activity detection strategies, the best TB classification performance is achieved after training using the dataset generated using XLS-R optimised for coverage.
This strategy exceeded the performance achieved by TB classification models trained using the dataset extracted by AST by 6\% absolute in development set AUC, and 4\% in test AUC (XLS-R (C) vs. AST (C)).
The TB classifier trained using the coughs automatically isolated by XLS-R is only marginally outperformed, by 2\% absolute in both development and test AUC, by the classifier trained on the manually annotated coughs.
Therefore the application of XLS-R to the task of automatic extraction of coughs from continuous audio appears a feasible approach to automated cough screening.

The application of median filtering is not found to consistently improve the quality of the isolated coughs such that downstream classification performance improves.
For both XLS-R and AST cough activity detection strategies, the AUC of the downstream TB classifiers is affected by up to only $1\%$.

\begin{table}[]
    \centering
    \caption{Ten fold nested cross-validation results for a downstream bidirectional LSTM TB classification model when trained using one of the nine automatically extracted cough datasets.
    C: coverage, EE: equal error rate, P: purity,
    \textdagger: After application of median filter.
    Manual: Classification performance when training on manual (ground truth) coughs.
    }
    \label{tab:results_lstm_classification_results}
    \vspace{2mm}
    \setlength{\tabcolsep}{4pt}
    \begin{tabular}{l l c c}
        \toprule
        Operating & \multirow{2}{*}{Model} & Development & Test \\
        point &  & AUC & AUC \\
        \midrule
                      & Manual             & $0.66 \pm 0.01$ & $0.65 \pm 0.07$ \\
        \midrule
        \multirow{2}{*}{C} & AST & $0.58 \pm 0.02$ & $0.59 \pm 0.09$ \\ 
                                & XLS-R     & $\bf{0.64 \pm 0.02}$ & $\bf{0.63 \pm 0.10}$ \\
        \cmidrule{2-4}
        \multirow{2}{*}{C$^\dagger$} & AST       & $0.59 \pm 0.02$ & $0.57 \pm 0.09$ \\
                                         & XLS-R     & $0.63 \pm 0.02$ & $0.62 \pm 0.09$ \\
        \midrule
        \multirow{3}{*}{EE} & LR        & $0.59 \pm 0.03$ & $0.59 \pm 0.10$ \\ 
                           & AST       & $0.60 \pm 0.01$ & $0.59 \pm 0.06$ \\
                                    & XLS-R     & $0.62 \pm 0.01$ & $0.61 \pm 0.10$ \\ 
        \cmidrule{2-4}
        \multirow{2}{*}{EE$^\dagger$} & AST       & $0.61 \pm 0.01$ & $0.60 \pm 0.05$ \\
                                         & XLS-R     & $0.62 \pm 0.02$ & $0.62 \pm 0.09$ \\
        \midrule
        \multirow{2}{*}{P} & AST   & $0.57 \pm 0.03$ & $0.57 \pm 0.12$ \\ 
                            & XLS-R         & $0.61 \pm 0.02$ & $0.61 \pm 0.10$ \\ 
        \bottomrule
    \end{tabular}
\end{table}

\section{Conclusions}
\label{sec:conclusion}

Three architectures have been investigated for the task of automatic cough activity detection as a part of the pipeline of a cough audio based TB screening system.
A large transformer-based network, XLS-R, was shown to outperform both AST and LR by 9\% and 27\% absolute in terms of test set average precision, despite having been pretrained only on speech.
Nine datasets were compiled by automatically isolating cough sounds using each of the cough detection models and respectively optimising the decision threshold for coverage, equal error rate, and purity.
When these datasets were used to train a downstream TB classification model, it was found that the coughs isolated utilising XLS-R afforded consistently the best test set AUC, outperforming AST by 4\% absolute in test set AUC, and were only narrowly outperformed (by 2\% absolute) by a classifier trained on the manually annotated coughs.

\bibliographystyle{IEEEbib}
\bibliography{refs}

\end{document}